\documentclass{Interspeech}

\interspeechcameraready

\usepackage{algorithmic}
\usepackage{bm}
\usepackage{subcaption}
\usepackage{comment}
\usepackage{cite}
\usepackage{multirow}
\usepackage{mathtools}

\title{OWSM-Biasing: Contextualizing Open Whisper-Style Speech Models \\for Automatic Speech Recognition with Dynamic Vocabulary}

\author[affiliation={1}]{Yui}{Sudo}
\author[affiliation={1}]{Yusuke}{Fujita}
\author[affiliation={1}]{Atsushi}{Kojima}
\author[affiliation={1}]{Tomoya}{Mizumoto}
\author[affiliation={1}]{Lianbo}{Liu}

\affiliation{}{SB Intuitions}{Japan}
\email{yui.sudo@sbintuitions.co.jp}

\usepackage{comment}

\begin{document}

\maketitle

\begin{abstract} % 1000 characters. ASCII characters only. No citations.
Speech foundation models (SFMs), such as Open Whisper-Style Speech Models (OWSM), are trained on massive datasets to achieve accurate automatic speech recognition. However, even SFMs struggle to accurately recognize rare and unseen words. While contextual biasing (CB) is a promising approach to improve recognition of such words, most CB methods are trained from scratch, resulting in lower performance than SFMs due to the lack of pre-trained knowledge. This paper integrates an existing CB method with OWSM v3.1 while freezing its pre-trained parameters. By leveraging the knowledge embedded in SFMs, the proposed method enables effective CB while preserving the advantages of SFMs, even with a small dataset. Experimental results show that the proposed method improves the biasing word error rate (B-WER) by 11.6 points, resulting in a 0.9 point improvement in the overall WER while reducing the real-time factor by 7.5\% compared to the non-biasing baseline on the LibriSpeech 100 test-clean set.
\end{abstract}

\section{Introduction}

Speech foundation models (SFMs) \cite{radford2022whisper,peng2023reproducing,google-usm,meta-mms,nv-canary,peng2024owsmctc,peng2024owsm}, such as OpenAI’s Whisper \cite{radford2022whisper} and Open Whisper-Style Speech Models (OWSM) \cite{peng2023reproducing}, have demonstrated remarkable performance in automatic speech recognition (ASR). These models are trained on massive datasets, allowing them to generalize well across multiple domains and achieve high recognition accuracy. Despite their strong performance, SFMs still struggle to recognize rare and unseen words due to their low occurrence in the training data.

Recent studies have shown that prompting SFMs with contextual words can improve recognition accuracy for these words \cite{li-etal-2024-cb,peng2024owsm}.
Surprisingly, SFMs exhibit some degree of contextualization ability when prompted with relevant words, despite the lack of explicit training for this purpose. Nonetheless, it has been reported that this emergent ability is not observed in smaller models \cite{peng2024owsm}. 
While similar approaches have been explored for Whisper \cite{sun23e_interspeech,monteiro24_interspeech,10800265,li24c_interspeech}, these prompt-based methods increase input token length, resulting in higher computational complexity.
In addition, speech language models (SpeechLMs) \cite{chu2023qwen,tang2024salmonn,ma2024embarrassingly,peng2024voicetextblender} have been explored for contextualization and error correction using prompts in ASR tasks \cite{gong24b_interspeech,higuchi2023harnessing}. Still, their performance degrades as the number of contextual words increases \cite{gong24b_interspeech}.

Contextual biasing (CB) has been actively explored to improve the recognition of rare and unseen words using an editable biasing list \cite{deepcontext2018,Jain2020ContextualRF,huber2021instant,sudo2023retraining,huang2023contextualized,zhou2023copyne}. Unlike prompt-based methods, CB methods are explicitly trained for this purpose, achieving higher accuracy on these words.
However, most existing CB methods are trained from scratch, often resulting in lower overall performance than SFMs due to the lack of large-scale training datasets. In addition, these methods typically require several hundred to thousands of hours of speech data, such as LibriSpeech-960 \cite{panayotov2015librispeech,shakeel2024_bias} and corpus of spontaneous Japanese (581 hours) \cite{csj,futami2023phoneme}. This makes them impractical for data-scarce domains.
Therefore, the integration of CB methods with SFMs is desirable to leverage large-scale knowledge while improving contextual word recognition. 

In practice, conventional CB methods have become increasingly complex to improve contextual word recognition. For example, several CB methods tightly couple biasing modules with the encoder or decoder, making integration with SFMs challenging \cite{nakagome24_interspeech,promptasr2024,sudo2024_bias,qiu2023improving,Le2021ContextualizedSE}.
Recently, the dynamic vocabulary-based CB method \cite{sudo2024contextualized} has been proposed as a simple yet highly effective alternative. This method dynamically expands the vocabulary by adding lightweight extensions to the embedding and output layers. 
This design can be applied to various end-to-end ASR architectures, such as connectionist temporal classification (CTC), attention-based encoder-decoder, recurrent neural network transducer (RNN-T), and their hybrids that perform joint decoding \cite{sudo2024contextualized,watanabe2017hybrid,sudo23c_interspeech,sudo20244d}, without modifying the core components. 
While this approach has shown promising results, its integration with SFMs remains unexplored. To the best of our knowledge, no prior work has attempted to integrate dynamic vocabulary-based CB methods into SFMs. It remains non-trivial to enable CB while preserving the strengths of SFMs, such as generalization ability and compatibility with pre-trained components.
In addition, most previous studies \cite{li-etal-2024-cb,sun23e_interspeech,qiu2023improving,Le2021ContextualizedSE,sudo2024contextualized} have not evaluated the computational cost despite introducing additional modules, raising concerns about its feasibility for large-scale SFMs.

This paper integrates the dynamic vocabulary-based CB method with SFMs to enable effective CB while preserving the strengths of SFMs.
Specifically, we combine OWSM v3.1 \cite{peng2024owsm} with the dynamic vocabulary-based CB method \cite{sudo2024contextualized}, while freezing OWSM v3.1 during training. 
Our method enables effective CB while leveraging the knowledge embedded in OWSM v3.1, reducing the need for large-scale datasets.
We also evaluate the computational complexity of the proposed method to ensure its feasibility for large-scale SFMs.
The main contributions of this paper are as follows:
\begin{itemize}
    \item We integrate OWSM v3.1 and the dynamic vocabulary-based CB method, enabling effective CB while preserving the advantages of the SFM, even with a small dataset.
    \item We show that the proposed method improves the recognition of rare and unseen words compared to OWSM v3.1 while reducing the real-time factor (RTF).
    \item We demonstrate that the proposed method remains effective across different OWSM versions.
\end{itemize}

\section{OWSM v3.1}
\label{sec:Preliminary}

This section provides an overview of OWSM v3.1 \cite{peng2024owsm}, which is integrated into the dynamic vocabulary-based CB method \cite{sudo2024contextualized} in Section \ref{sec:proposed}.

\subsection{OWSM encoder}
\label{sec:encoder}

OWSM v3.1 adopts the E-Branchformer \cite{e-branchformer} for the encoder.
The OWSM encoder consists of stacked E-Branchformer layers, which utilize parallel branches to capture both local and global features. %, merging them through convolutional layers.
The encoder transforms an audio feature sequence \begin{math}\bm{X}\end{math} into a $T$-length $d$-dimensional hidden representation $\bm{H}$ as follows:
\begin{equation}
\label{eq:audio}
    \bm{H} = \mathrm{OwsmEncoder}(\bm{X}) \in \mathbb{R}^{T \times d}.
\end{equation}

\subsection{OWSM decoder}
\label{sec:attention}

The OWSM decoder consists of an embedding layer, Transformer layers, and an output layer.
Given $\bm{H}$ %generated by the OWSM encoder 
in Eq.~\eqref{eq:audio} and the previous token sequence $y_{0:i-1}$, the OWSM decoder estimates the $i$-th token $y_i \in \mathcal{V}^{\text{s}}$, where $\mathcal{V}^{\text{s}}$ represents the pre-defined vocabulary of size $K$.

First, the embedding layer converts the previous token sequence $y_{0:i-1}$ to an embedding vector sequence $\bm{E}_{0:i-1}$ as follows:
\begin{equation}
    \bm{E}_{0:i-1} = \mathrm{Embedding}(y_{0:i-1}) \in \mathbb{R}^{i \times d}.
\label{eq:embedding}
\end{equation}
Then, $\bm{E}_{0:i-1}$ is fed into the OWSM decoder, along with the hidden representation $\bm{H}$ in Eq.~\eqref{eq:audio}, to generate the $i$-th hidden state vector $\bm{u}_i$ as follows:
\begin{equation}
    \bm{u}_{i} = \mathrm{OwsmDecoder}(\bm{H}, \bm{E}_{0:i-1}) \in \mathbb{R}^{d}.
\label{eq:main_decoder}
\end{equation}
Finally, the output layer computes the token-wise score $\bm{\alpha}^{\text{s}}_{i}$ and the corresponding probability distribution $\bm{z}_{i}$ as follows:
\vspace{-0.5mm}
\begin{align}
\label{eq:score}
    & \bm{\alpha}^{\text{s}}_{i} = \mathrm{Linear}(\bm{u}_{i}) \in \mathbb{R}^{K},\\
\label{eq:normal_softmax}
    & \bm{z}_{i} = \mathrm{Softmax}(\bm{\alpha}^{\text{s}}_i) \in \mathbb{R}^{K}.
\end{align}
By recursively applying this process, the OWSM decoder generates the $I$-length token-wise probability distribution $\bm{Z} \in \mathbb{R}^{I \times K}$ of the token sequence $Y \in \mathbb{R}^{I}$.

\section{Integration with dynamic vocabulary}
\label{sec:proposed}

\begin{figure}[t]
        \begin{minipage}{0.49\textwidth}
        \centering
            \includegraphics[width=0.92\textwidth]{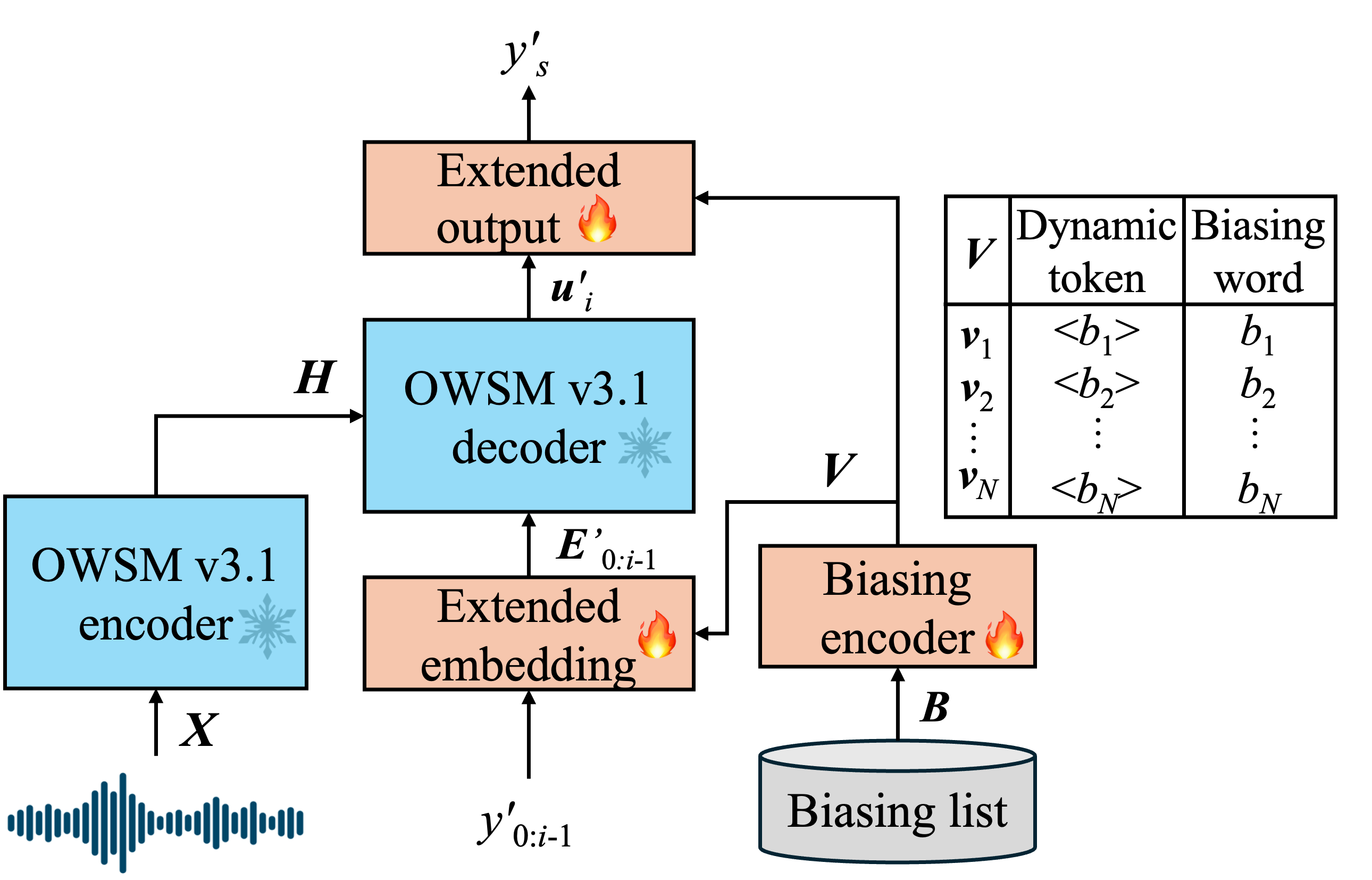} 
        \end{minipage}
    \vspace*{-2mm}
    \caption{Overall architecture of the proposed method, which integrates OWSM v3.1 \cite{peng2024owsmctc} and the dynamic vocabulary-based CB method \cite{sudo2024contextualized}. Red and blue block represent trainable and frozen modules, respectively. %The parameters of the trainable modules are much smaller than the frozen modules.
    }
    \label{fig:overview}
\vspace*{-3mm}
\end{figure}

Figure \ref{fig:overview} shows the overall architecture of the proposed method, which integrates OWSM v3.1 with the dynamic vocabulary-based CB method \cite{sudo2024contextualized}. 
Since this method does not change the core structures of the encoder and decoder, only the biasing modules (Figure \ref{fig:components}) are added to OWSM v3.1 while freezing the pre-trained parameters.
The following subsections describe each module and the training methodology.

\subsection{Biasing encoder}
\label{sec:biasencoder}

\begin{figure*}[t]
     \centering
     \hfill
     \begin{subfigure}[b]{0.45\linewidth}
         \centering
         \includegraphics[scale=0.4]{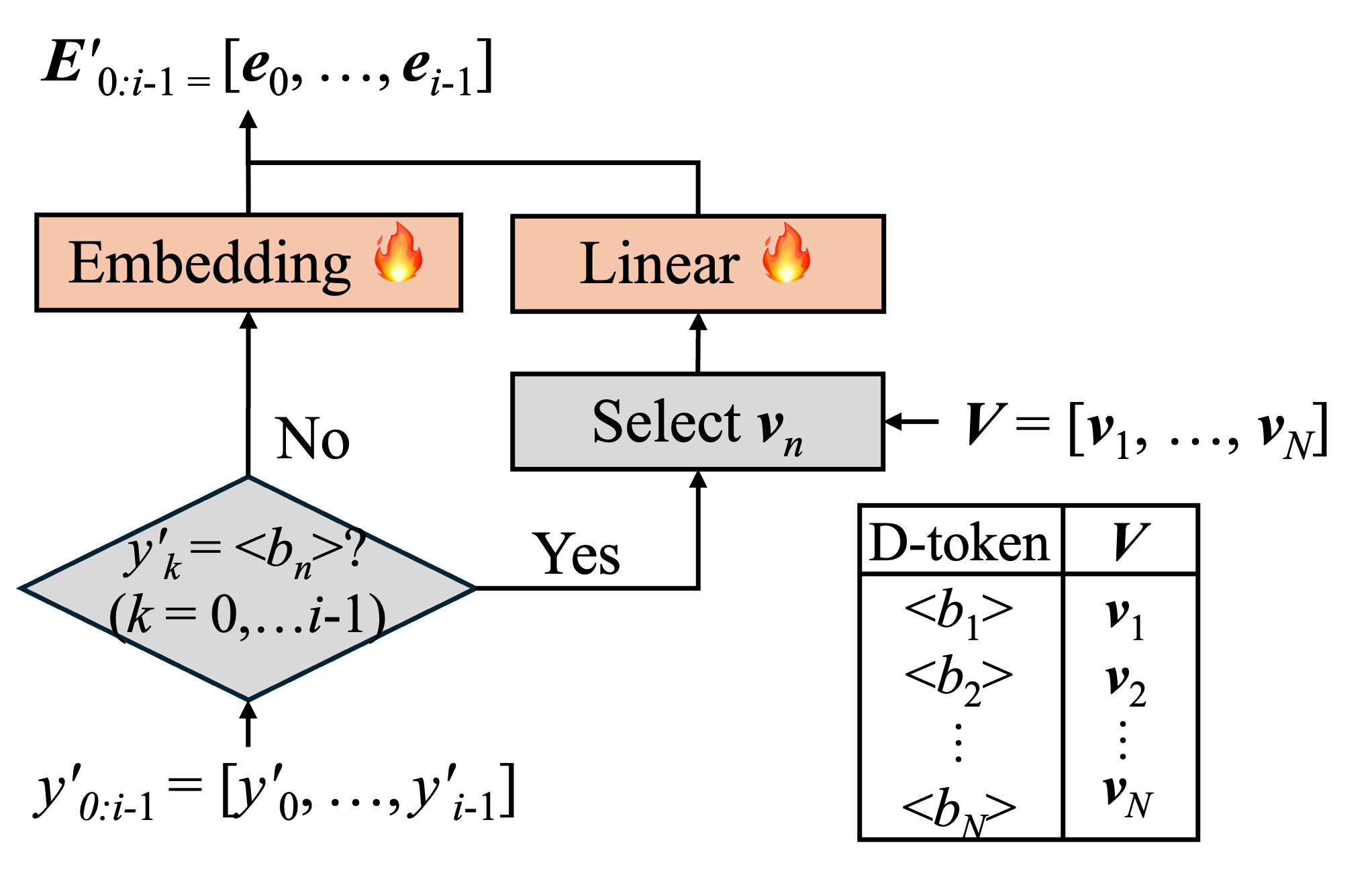}
         \vskip -2.0mm
         \caption{Extended embedding layer}
         \label{fig:embeding}
     \end{subfigure}
     \hfill
     \begin{subfigure}[b]{0.27\linewidth}
         \centering
         \hskip 2.0mm
         \includegraphics[scale=0.4]{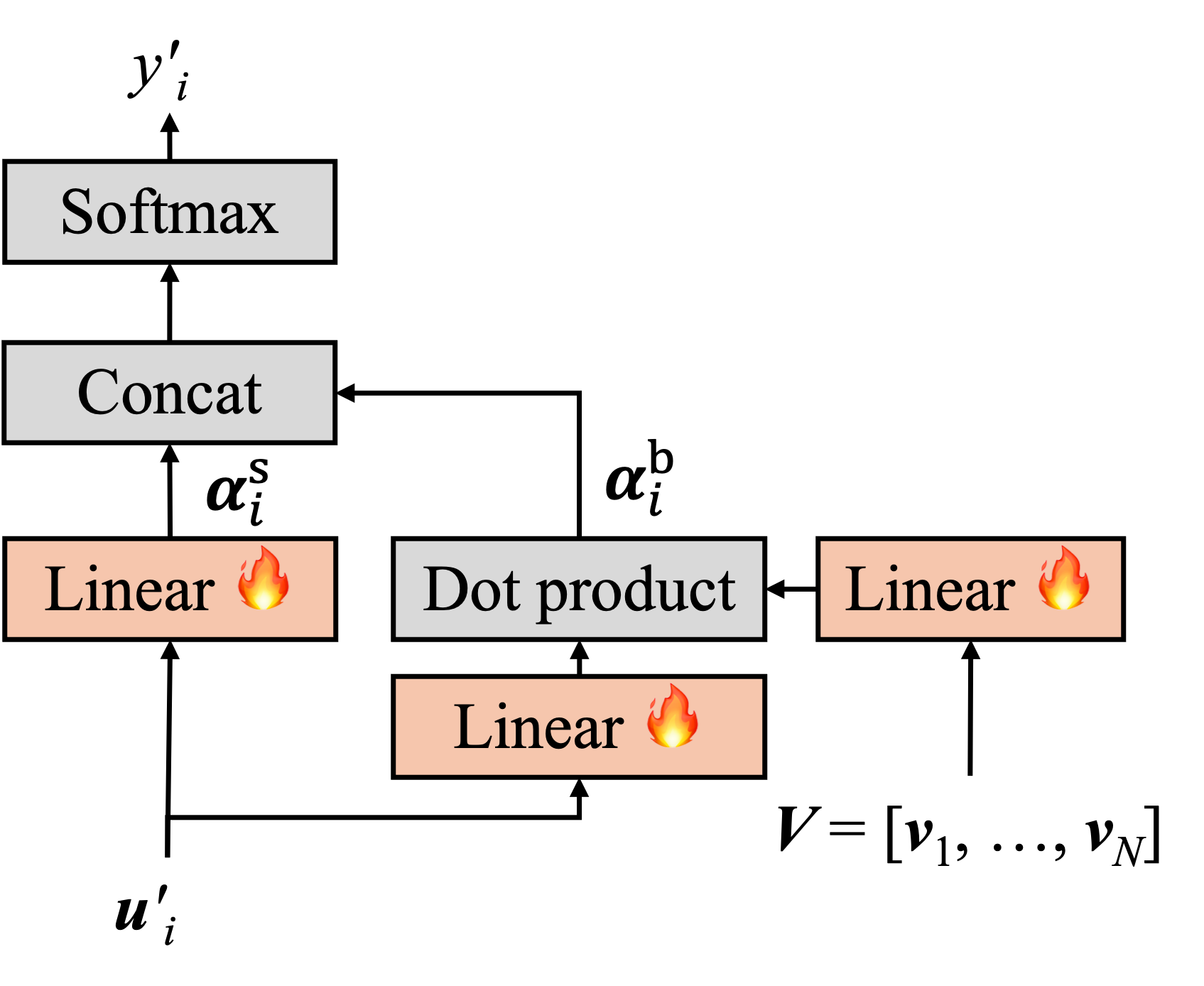}
         \vskip -2.0mm
         \caption{Extended output layer}
         \label{fig:output}
     \end{subfigure}
     \hfill
     \begin{subfigure}[b]{0.27\linewidth}
         \centering
         \hskip 2.0mm
         \includegraphics[scale=0.4]{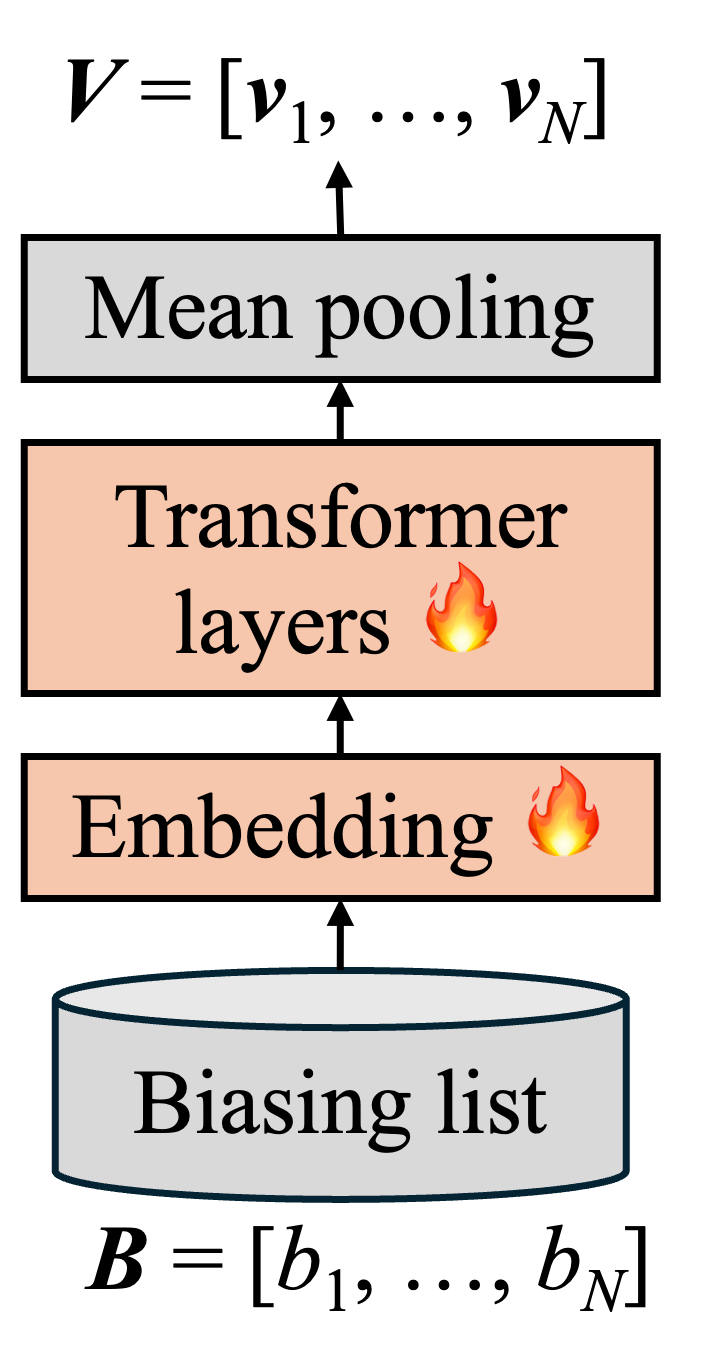}
         \vskip -2.0mm
         \caption{ Biasing encoder}
         \label{fig:bias-encoder}
     \end{subfigure}
     \hfill
    \vskip -6.0mm
    \caption{The biasing modules of the dynamic vocabulary. Red and gray blocks show trainable and non-trainable components.}
    \label{fig:components}
    \vskip -3mm
\end{figure*}

The biasing encoder consists of Transformer \cite{NIPS2017_3f5ee243} layers and a mean pooling (Figure \ref{fig:bias-encoder}). 
The biasing encoder extracts the $d^{\prime}$-dimensional embeddings $\bm{V} = [\bm{v}_1, \cdots , \bm{v}_{N}]$ from a biasing list $\bm{B} = \{b_{1}, \cdots, b_{N}$\}, where $b_{n}$ represents a biasing word, as follows:
\begin{equation}
    \bm{V} = \mathrm{BiasingEnc}(\bm{B}) \in \mathbb{R}^{N \times d^{\prime}}.
\label{eq:hp}
\end{equation}
Each biasing word $b_{n}$ is represented as a sequence of subword tokens from $\mathcal{V}^{\text{s}}$ (e.g., [``\textit{all}'', ``\textit{ig}'', ``\textit{at}'',``\textit{or}'']) and is dynamically incorporated into the dynamic vocabulary $\mathcal{V}^{\text{b}} = \{<\hspace{-3pt}b_1\hspace{-3pt}>, \cdots, <\hspace{-3pt}b_N\hspace{-3pt}>\}$, where each entry serves as a single dynamic token representing the corresponding word (e.g., [$<$\textit{alligator}$>$]).

\subsection{Decoder with extended embedding and output layers}
\label{sec:biasdecoder}

Unlike the conventional OWSM v3.1, the decoder with the extended embedding and output layers estimates the next token $y_i^{\prime}$ from the extended vocabulary $\mathcal{V}^{\text{s}} \cup \mathcal{V}^{\text{b}}$. %, based on $\bm{H}$, $\bm{V}$, and $y^{\prime}_{0:i-1}$. 
For example, if the biasing word ``\textit{alligator}'' exists in the biasing list, the decoder outputs the single dynamic token $<$\textit{alligator}$>$ in $\mathcal{V}^{\text{b}}$ instead of a sequence of static tokens [``\textit{all}'', ``\textit{ig}'', ``\textit{at}'',``\textit{or}''] in $\mathcal{V}^{\text{s}}$.

First, each token in $y^{\prime}_{0:i-1}$ is converted into an embedding vector, resulting in the sequence 
$\bm{E}^{\prime}_{0:i-1} = [\bm{e}^{\prime}_0, \cdots , \bm{e}^{\prime}_{i-1}] \in \mathbb{R}^{d \times i}$.
Unlike Eq.~\eqref{eq:embedding}, if $y^{\prime}_{i-1}$ is a dynamic token in $\mathcal{V}^{\text{b}}$ 
($y^{\prime}_{i-1} = <\hspace{-3pt}b_n\hspace{-3pt}>$), its corresponding embedding $\bm{v}_{n}$ is selected from $\bm{V}$ using $\mathrm{Select}(\cdot)$; otherwise, $\mathrm{Embedding}(\cdot)$ in Eq.~\eqref{eq:embedding} is applied, as follows:
\begin{equation}
\label{eq:extended_embed}
   \bm{e}^{\prime}_{i-1} = \begin{cases}
                    \mathrm{Embedding}(y^{\prime}_{i-1}) & (y^{\prime}_{i-1} \in \mathcal{V}^{\text{s}})\\
                    \mathrm{Linear}(\mathrm{Select}(\bm{V}, y^{\prime}_{i-1})) & (y^{\prime}_{i-1} \in \mathcal{V}^{\text{b}}).
                \end{cases}
\end{equation}
For example, if $y^{\prime}_{i-1} = <\hspace{-3pt}b_3\hspace{-3pt}>$, $\mathrm{Select}(\cdot)$ returns $\bm{v}_3 \in \mathbb{R}^{d^{\prime}}$, which is the third row of $\bm{V} = [\bm{v}_1, \cdots, \bm{v}_{N}]$, and is then transformed into $\bm{e}^{\prime}_{i-1} \in \mathbb{R}^{d}$ through a linear layer.
Here, we slightly modify the extended embedding layer proposed in \cite{sudo2024contextualized} by repositioning the linear layer to avoid altering the embeddings of static vocabulary (Figure \ref{fig:embeding}).

The OWSM decoder then converts $\bm{E}^{\prime}_{0:i-1}$ into the hidden state vector $\bm{u}^{\prime}_{i}$ as in Eq. \eqref{eq:main_decoder}: 
\begin{equation}
    \bm{u}^{\prime}_{i} = \mathrm{OwsmDecoder}(\bm{H}, \bm{E}^{\prime}_{0:i-1}) \in \mathbb{R}^{d}.
\end{equation}
Finally, the extended output layer computes the similarity score $\bm{\alpha}^{\text{b}}_i$ using a dot product for dynamic tokens and combines it with $\bm{\alpha}^{\text{s}}_i$, in contrast to Eq. \eqref{eq:normal_softmax}, as follows (Figure \ref{fig:output}):
\begin{align}
    & \bm{\alpha}^{\text{s}}_i = \mathrm{Linear}(\bm{u}_{i}^{\prime})\in \mathbb{R}^{K}, \\
    \label{eq:inner}
    & \bm{\alpha}^{\text{b}}_i = \frac{\mathrm{Linear}(\bm{u}^{\prime}_{i}) \mathrm{Linear}(\bm{V}^{T})}{\sqrt{d}} \in \mathbb{R}^{N}, \\
    & \bm{z}^{\prime}_i = \mathrm{Softmax}(\mathrm{Concat}(\bm{\alpha}^{\text{s}}_i, \bm{\alpha}^{\text{b}}_i)) \in \mathbb{R}^{K + N},
\end{align}
where $\bm{z}^{\prime}_{i}$ represents the token-wise probability distribution over both static and dynamic vocabulary.
By recursively applying this process, the $I^{\prime}$-length token-wise probability distribution $\bm{Z}^{\prime} \in \mathbb{R}^{I^{\prime} \times (K + N)}$ of the token sequence $Y^{\prime} \in \mathbb{R}^{I^{\prime}}$ is obtained.

As described above, the dynamic vocabulary-based CB method does not change the core structures of the encoder and decoder, allowing the method to be easily integrated with various SFMs, such as \cite{radford2022whisper,peng2024owsm,nv-canary}.

\subsection{Training and decoding}
\label{sec:training}

To preserve the advantages of OWSM v3.1, we freeze the parameters of the OWSM encoder and decoder as shown in Figure \ref{fig:overview}.
Figure \ref{fig:components} illustrates the trainable components in the biasing modules, where red and gray blocks represent the trainable and non-trainable components, respectively. 
We train the embedding and output layers of OWSM with vocabulary size $K$ of 5,000.
The trainable modules primarily consist of lightweight components, such as linear and embedding layers. Although the biasing encoder employs Transformer layers, which are computationally expensive, it remains smaller than the frozen modules of OWSM v3.1 (discussed further in Section \ref{sec:experimental condition}).

For training and decoding settings, we follow the same strategy as \cite{sudo2024contextualized}.
During training, a biasing list $\bm{B}$ containing $N$ biasing words is randomly generated from the reference transcriptions for each batch.
During decoding, we apply the biasing weight $\mu$ to adjust the token-wise probability to mitigate over-biasing and under-biasing.

\section{Experiment}

We conduct several experiments to verify the effectiveness of the proposed method.

\subsection{Experimental setup}
\label{sec:experimental condition}

Table~\ref{condition} shows the model configurations of the proposed method, which consists of OWSM v3.1 and the biasing modules. OWSM v3.1 employs the E-Branchformer \cite{e-branchformer} and Transformer as the encoder and decoder, respectively. We use both the base and medium versions of OWSM v3.1 in our experiments.
The biasing encoder is identically configured for both versions, accounting for 26.3\% and 3.1\% of the total parameters, respectively.
The base and medium models are trained for 100 and 200 epochs, respectively, with a learning rate of 0.002 and 15,000 warmup steps using the Adam optimizer \cite{KingmaB14}.
The biasing weight $\mu$ (Section \ref{sec:training}) is set to 0.3.

The proposed method is evaluated on the LibriSpeech 100 corpus \cite{panayotov2015librispeech} assuming a data-scarce domain in terms of word error rate (WER) and biasing WER (B-WER) \cite{Le2021ContextualizedSE} using the ESPnet toolkit \cite{espnet}. 
We use the same biasing list as in \cite{Le2021ContextualizedSE}, which is constructed based on word frequency in the training data.
We also measure RTF using an NVIDIA A100 GPU to assess the computational cost, with a beam size of 3, with a beam size of 3 using joint CTC/attention beam search \cite{watanabe2017hybrid,sudo20244d}.
Note that the biasing encoder is excluded from the RTF measurement, because the resulting embedding $\bm{V}$ in Eq. \eqref{eq:hp} can be retained unless the biasing list is updated, which is infrequently (e.g., not on a per-utterance basis).

\begin{table}[t]
\caption{Model configurations. The proposed method adds the biasing modules to OWSM v3.1.}
\vspace{-7mm}
\label{condition}
\begin{center}
\resizebox {0.78\linewidth} {!} {
\begin{tabular}{@{}lcc}
\toprule
 & \multicolumn{2}{c}{\textbf{OWSM-Biasing}} \\
   & Base & Medium \\
\midrule
\multicolumn{3}{l}{\textbf{OWSM v3.1} (shown in blue in Figure \ref{fig:overview})} \\
Encoder & \multicolumn{2}{c}{E-Branchformer} \\
Decoder & \multicolumn{2}{c}{Transformer} \\
Layers & 6 & 18 \\
Hidden & 384 & 1024 \\
Heads & 6 & 16 \\
Linear units & 1536 & 4096 \\
%Params (frozen) & 101M & 1.02B \\
Params (frozen) & 59.0M & 890.6M \\
\midrule
\multicolumn{3}{l}{\textbf{Biasing module} (shown in red in Figure \ref{fig:overview})} \\
Biasing encoder (Figure \ref{fig:bias-encoder}) & \multicolumn{2}{c}{Transformer} \\
Layers & \multicolumn{2}{c}{6} \\
Hidden & \multicolumn{2}{c}{256} \\
Heads & \multicolumn{2}{c}{4} \\
Embedding (Figure \ref{fig:embeding}) & \multicolumn{2}{c}{Embedding, linear}\\
Output (Figure \ref{fig:output}) & \multicolumn{2}{c}{Linear $\times$ 3}\\
Params (trainable) & 15.5M & 27.4M \\
\midrule
\multicolumn{3}{l}{\textbf{Total params}} \\
Total params & 74.5M & 928.0M \\
Trainable rate & 26.3\% & 3.1\% \\
\bottomrule
\end{tabular}
}
\end{center}
\vspace*{-8mm}
\end{table}

\begin{table*}[t]
\caption{WER (B-WER) comparison on LibriSpeech 100. \textbf{Bolded} words represent the best result among the same biasing list size $N$.}
\vspace*{-7mm}
\label{maintable}
\begin{center}
\resizebox {\linewidth} {!} {
\begin{tabular}{@{}c|l|cccc|cccc}
\toprule
&      & \multicolumn{4}{c|}{test-clean} & \multicolumn{4}{c}{test-other} \\
ID & Model & $N$=100 & $N$=500 & $N$=1000 & $N$=2000 & $N$=100 & $N$=500 & $N$=1000 & $N$=2000\\
\midrule
A1 & OWSM v3.1 base \cite{peng2024owsm} & \multicolumn{4}{c|}{3.9 (15.5)} & \multicolumn{4}{c}{9.5 (32.2)} \\
A2 & Dynamic vocabulary (scratch) \cite{sudo2024contextualized} & 4.7 (8.5) & 4.9 (9.2) & 5.0 (9.7) & 5.4 (10.4) & 15.0 (23.3)  & 15.4 (24.5) & 15.9 (26.1) & 16.5 (27.9) \\
A3 & + OWSM v3.1 base initialization & 3.6 (6.8) & 3.8 (7.2) & 4.0 (7.8) & 4.3 (8.8) & 11.5 (18.1) & 11.8 (19.4)  & 12.3 (20.6) & 12.6 (22.3)\\
A4 & \textbf{OWSM-Biasing base (ours)} & \textbf{3.0} (\textbf{3.9}) & \textbf{3.2} (\textbf{4.9}) & \textbf{3.4} (\textbf{5.7}) & \textbf{3.8} (\textbf{6.9}) & \textbf{7.9} (\textbf{11.5})& \textbf{8.3} (\textbf{13.0}) & \textbf{8.7} (\textbf{14.5}) & \textbf{9.2} (\textbf{16.7}) \\ 
\bottomrule
\end{tabular}
}
\end{center}
\vspace*{-3mm}
\end{table*}

\begin{table*}[t]
\caption{Decoding examples. 
%in the LibriSpeech test-other set (utterance ID: 2414-159411-0018).
\textbf{Bolded} words represent the biasing words, and \textcolor{red}{red} faces represent incorrectly recognized words.}
\vspace*{-7mm}
\label{example}
\begin{center}
\resizebox {\linewidth} {!} {
\begin{tabular}{@{}c|l|l}
\toprule
ID &Reference & after this they saw an \textbf{alligator} and the \textbf{brahman} \textbf{related} the matter to him hoping for a more favorable \textbf{verdict}  \\
\midrule
A1 & OWSM v3.1 base \cite{peng2024owsm} & after this they saw an \textcolor{red}{\textbf{allicator}} and the \textcolor{red}{\textbf{brammel}} \textbf{related} the matter to him hoping for a more favorable \textcolor{red}{\textbf{word it}} \\
A2 & Dynamic vocabulary (scratch) \cite{sudo2024contextualized} & after this they saw \textcolor{red}{and} \textbf{alligator} and the \textcolor{red}{\textbf{pegramer}} \textcolor{red}{\textbf{relearned}} the \textcolor{red}{mat} to him \textcolor{red}{who being} for a \textcolor{red}{most} favorable \textbf{verdict} \\
A3 & + OWSM v3.1 base initialization & after this \textcolor{red}{the} saw an \textbf{alligator} and the \textcolor{red}{\textbf{brown}} \textcolor{red}{\textbf{relearned}} the \textcolor{red}{mat} to him hoping for a more favorable \textbf{verdict} \\
A4 & \textbf{OWSM-Biasing base (ours)} & after this they saw an \textbf{alligator} and the \textbf{brahman} \textcolor{red}{\textbf{relearned}} the matter to him hoping for a more favorable \textbf{verdict}\\
\bottomrule
\end{tabular}
}
\end{center}
\vspace*{-6mm}
\end{table*}

\subsection{Main results}
\label{sec:main results}

Table \ref{maintable} and Table \ref{example} compare the performance of four models: OWSM v3.1 base \cite{peng2024owsm}, the dynamic vocabulary-based CB method trained from scratch \cite{sudo2024contextualized}, the fine-tuned OWSM v3.1 for dynamic vocabulary-based CB, and the proposed method. In Table \ref{example}, bolded words indicate biasing words, while misrecognized words are shown in red.

While OWSM v3.1 has a strong overall WER, it still struggles to accurately recognize rare and unseen words, resulting in a degraded B-WER. In contrast, the dynamic vocabulary-based CB method significantly improves B-WER but lacks the extensive knowledge embedded in OWSM v3.1, resulting in a worse overall WER (A1 vs. A2). 
Although fine-tuning all parameters of OWSM v3.1 for the dynamic vocabulary-based CB method improves both overall WER and B-WER compared to the scratch training, this approach suffers from catastrophic forgetting, potentially degrading overall WER compared to OWSM v3.1 (A1 vs. A3). 
In contrast, the proposed method, which freezes the OWSM encoder and decoder, consistently achieves substantial improvements in both overall WER and B-WER, even with a small dataset (A4).
Although B-WER degrades as the biasing list size $N$ increases due to the inclusion of more phonetically similar words, the proposed method still consistently outperforms the baselines.

\subsection{Analysis on rare and unseen words}
\label{sec:frequency}

\begin{figure}[t!]
     \centering
     \hfill
     \begin{subfigure}[b]{0.49\linewidth}
         \centering
         \includegraphics[scale=0.35]{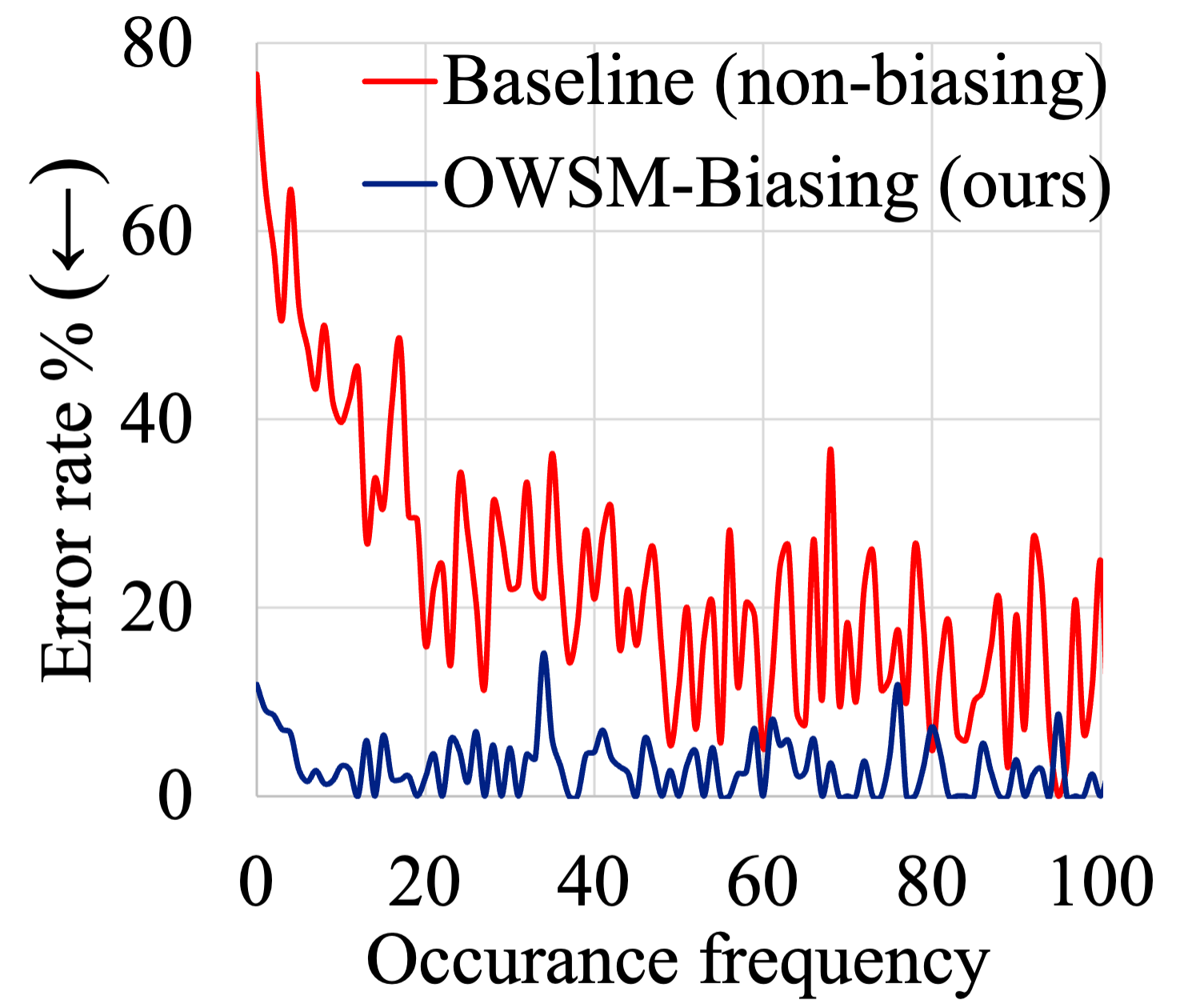}
         \vskip -2.0mm
         \caption{Error rate for rare words}
         \label{fig:freq}
     \end{subfigure}
     \hfill
     \begin{subfigure}[b]{0.49\linewidth}
         \centering
         \includegraphics[scale=0.33]{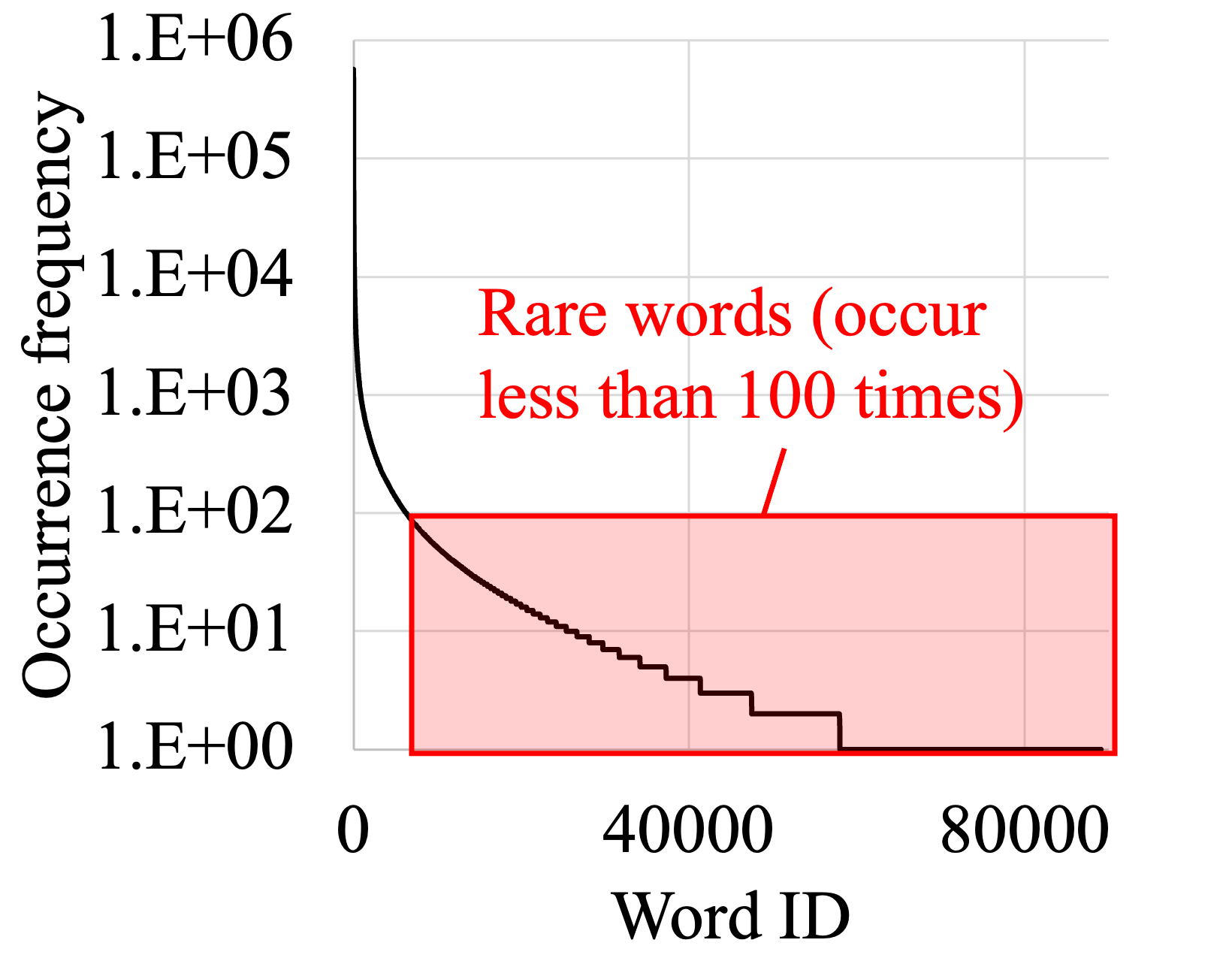}
         \vskip -2.0mm
         \caption{Frequency in training data}
         \label{fig:freq_training}
     \end{subfigure}
     \hfill
    \vskip -3.0mm
    \caption{Performance comparison on rare and unseen words.}
    \label{fig:frequency}
    \vskip -4.0mm
\end{figure}

Figure \ref{fig:freq} illustrates the error rates for rare and unseen words that occur less than 100 times in the training data. The red and blue lines represent the non-biasing baseline and the proposed method with a biasing list size of $N = 100$, respectively. The baseline struggles to accurately recognize these words, especially those that appear less than 50 times in the training data. In contrast, the proposed method consistently improves the error rate, even for completely unseen words.
Figure \ref{fig:freq_training} shows the words contained in the training data and their frequency of occurrence. While some common words, such as ``\textit{the}'' and ``\textit{and}'', occur more than 100,000 times, 90\% of the words occur less than 100 times in the training data. These results suggest that the proposed method remains effective, even as the training dataset grows larger.

\subsection{Comparison to prompt-based CB}
\label{sec:zero-shot}

Figure \ref{zero-shot} compares the proposed method with the prompt-based CB method \cite{peng2024owsm} for both OWSM v3.1 base and medium models. For reference, we also include the results of SpeechLM-based methods (C1, C2) \cite{chu2023qwen,gong24b_interspeech}.
While OWSM v3.1 medium (0.9B parameters) improves the B-WER by prompting, the base model shows little improvement. In contrast, the proposed method consistently improves the B-WER for both base and medium models (A1, B1 vs. A4, B4).
Although the overall WER of the proposed method (medium) is slightly worse than that of the prompt-based CB method (B1 vs. B4), the proposed method achieves a better balance between overall WER and B-WER, which is crucial for recognizing key contextual words, such as personal names and technical terms.

Larger models, such as SpeechLMs \cite{chu2023qwen,gong24b_interspeech} with over 7.0B parameters, further improve the overall WER. However, their impact on B-WER remains limited (C1 vs. C2). These results highlight the importance of the proposed method for improving contextual word recognition in large-scale models.

\begin{table}[t]
\caption{WER ($\downarrow$) comaprison to prompt-based CB methods.}
\vspace*{-7mm}
\label{zero-shot}
\begin{center}
\resizebox {0.99\linewidth} {!} {
\begin{tabular}{@{}c|l|cc|cc|c}
\toprule
& & \multicolumn{2}{c|}{test-clean} & \multicolumn{2}{c|}{test-other} & B-WER \\
ID & Model  & WER & B-WER & WER & B-WER & impr. ($\uparrow$) \\
\midrule
A1 & OWSM v3.1 base & 3.9 & 15.5 & 9.5 & 32.2 & -\\
A1 & + Prompt biasing \cite{peng2024owsm} & 4.4 & 14.8 & 12.5 & 30.4 & 5\% \\
A4 & \textbf{OWSM-Biasing (75M)} & \textbf{3.0} & \textbf{3.9} & \textbf{7.9} & \textbf{11.5} & \textbf{68\%} \\ %mu0.3
\midrule
B1 & OWSM v3.1 medium & 2.6 & 10.6 & 5.3 & 21.1 & -\\
B1 & + Prompt biasing \cite{peng2024owsm} & \textbf{2.2} & 7.3 & \textbf{5.0} & 15.4 & 21\% \\
B4 & \textbf{OWSM-Biasing (0.9B)} & 2.6 & \textbf{2.3} & 5.3 & \textbf{6.5} & \textbf{72\%} \\ 
\midrule
C1 & Qwen-Audio \cite{chu2023qwen} & 2.0 & 8.4 & 4.2 & 18.4 & -\\
C2 & + Biasing (7.6B) \cite{gong24b_interspeech} & 1.9 & 6.8 & 3.8 & 15.0 & 19\% \\
%C3 & + Biasing (7.6B) & 1.3 & 3.7 & 2.7 & 8.0 & \% \\
\bottomrule
\end{tabular}
}
\end{center}
\vspace*{-3mm}
\end{table}

\subsection{Computational complexity}
\label{sec:rtf}

\begin{figure}[t!]
     \centering
     \hfill
     \begin{subfigure}[b]{0.49\linewidth}
         \centering
         \includegraphics[scale=0.3]{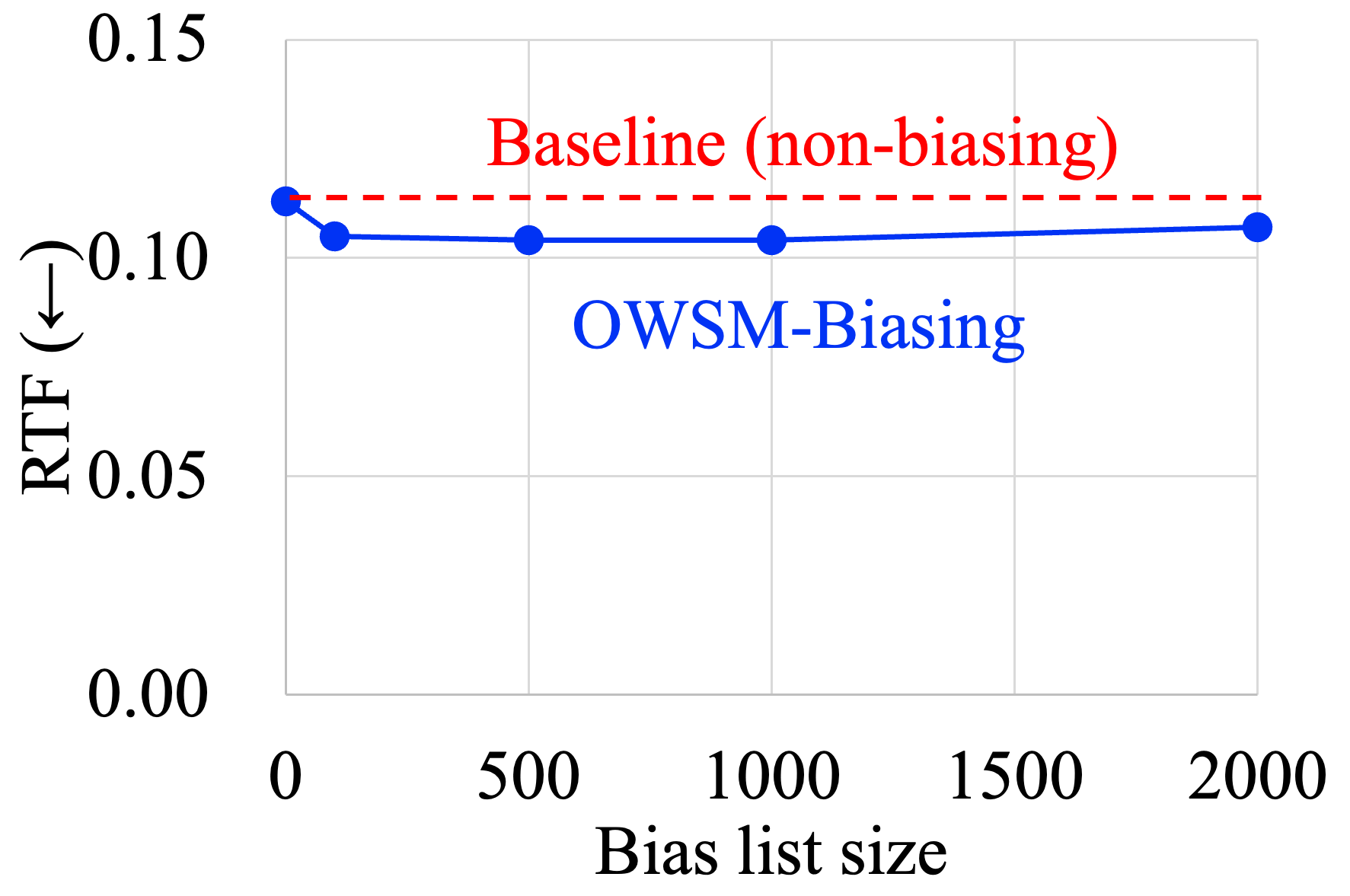}
         \vskip -2.0mm
         \caption{RTF}
         \label{fig:rtf}
     \end{subfigure}
     \hfill
     \begin{subfigure}[b]{0.49\linewidth}
         \centering
         \includegraphics[scale=0.3]{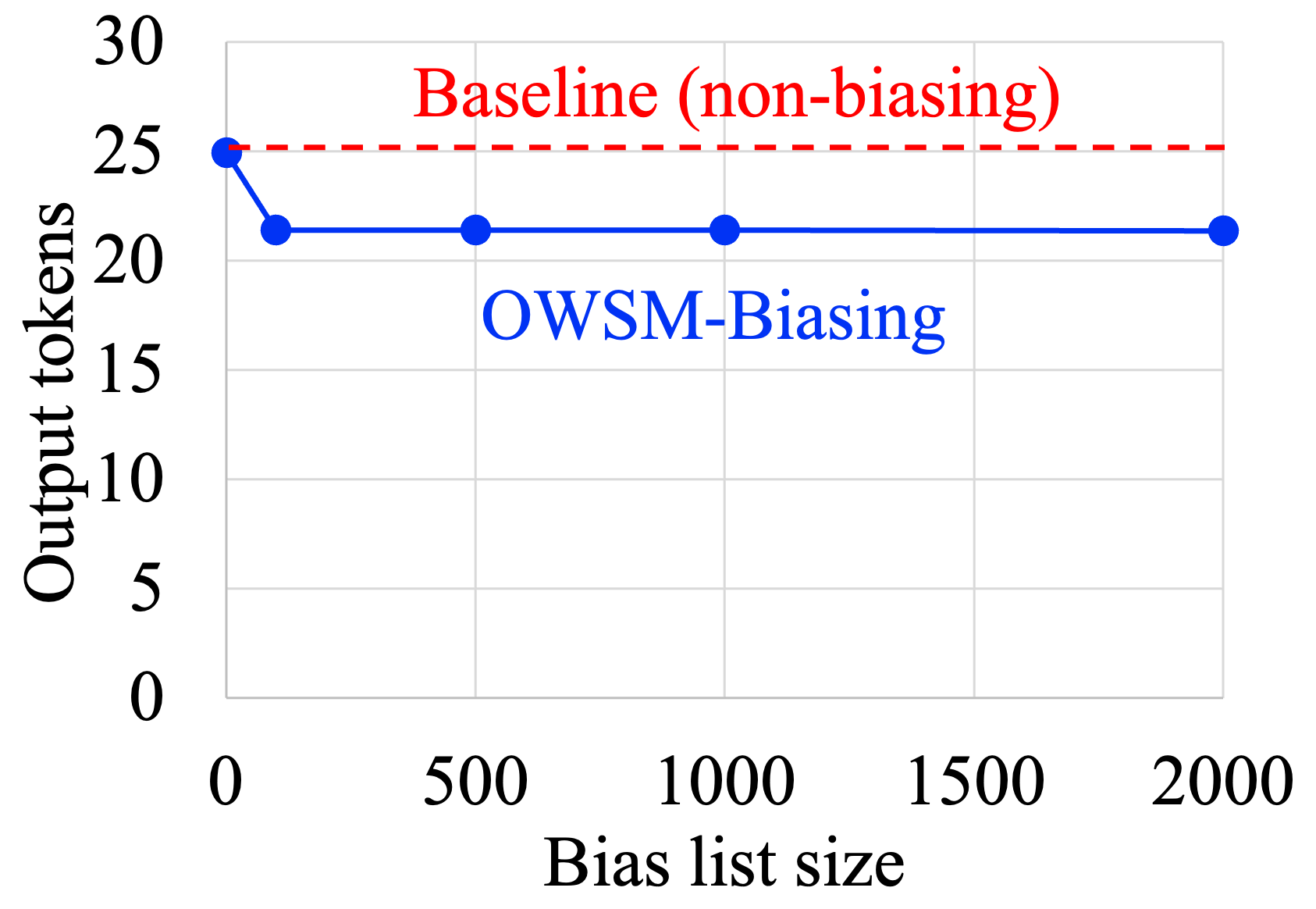}
         \vskip -2.0mm
         \caption{Number of decoding iterations}
         \label{fig:iteration}
     \end{subfigure}
     \hfill
    \vskip -3.0mm
    \caption{Computational complexity of the proposed method.}
    \label{fig:computational}
    \vskip -4.0mm
\end{figure}

Figure \ref{fig:rtf} compares the RTF between the non-biasing baseline and the proposed method. While the proposed method extends the embedding and output layers for CB, slightly increasing computational overhead with larger biasing lists, it reduces the RTF by 7.5\% on average. This is due to a 14.5\% reduction in decoding iterations, as shown in Figure \ref{fig:iteration}. Specifically, the proposed method outputs the biasing words as single dynamic tokens, eliminating the need to generate multiple subword tokens. For example, instead of generating [``\textit{all}'', ``\textit{ig}'', ``\textit{at}'',``\textit{or}''], the model directly outputs $<$\textit{alligator}$>$.
In contrast, the prompt-based CB \cite{peng2024owsm} inputs all biasing words to the decoder, leading to an $O(n^2)$ increase in computation due to the decoder's attention mechanisms. Given an average utterance length of 25 tokens (Figure \ref{fig:iteration}), adding 2,000 words as a prompt imposes excessive computational overhead. 

\section{Conclusion}

This paper integrates the dynamic vocabulary-based CB method with OWSM v3.1, freezing the pre-trained parameters to enable effective CB even with a small dataset, while preserving the advantages of SFMs. 
The proposed method improves the B-WER by 11.6 points, resulting in a 0.9 point improvement in the overall WER while reducing the RTF by 7.5\% compared to OWSM v3.1 on the LibriSpeech 100 test-clean set.

\clearpage

\bibliographystyle{IEEEtran}
\bibliography{mybib}

\begin{thebibliography}{10}
\providecommand{\url}[1]{#1}
\csname url@samestyle\endcsname
\providecommand{\newblock}{\relax}
\providecommand{\bibinfo}[2]{#2}
\providecommand{\BIBentrySTDinterwordspacing}{\spaceskip=0pt\relax}
\providecommand{\BIBentryALTinterwordstretchfactor}{4}
\providecommand{\BIBentryALTinterwordspacing}{\spaceskip=\fontdimen2\font plus
\BIBentryALTinterwordstretchfactor\fontdimen3\font minus \fontdimen4\font\relax}
\providecommand{\BIBforeignlanguage}[2]{{%
\expandafter\ifx\csname l@#1\endcsname\relax
\typeout{** WARNING: IEEEtran.bst: No hyphenation pattern has been}%
\typeout{** loaded for the language `#1'. Using the pattern for}%
\typeout{** the default language instead.}%
\else
\language=\csname l@#1\endcsname
\fi
#2}}
\providecommand{\BIBdecl}{\relax}
\BIBdecl

\bibitem{radford2022whisper}
R.~Alec, K.~Jong~W., X.~T., B.~G., M.~Christine, and S.~Ilya, ``Robust speech recognition via large-scale weak supervision.'' 2022.

\bibitem{peng2023reproducing}
Y.~Peng, J.~Tian, B.~Yan, D.~Berrebbi, X.~Chang \emph{et~al.}, ``Reproducing {Whisper}-style training using an open-source toolkit and publicly available data,'' in \emph{Proc. ASRU}, 2023, pp. 1--8.

\bibitem{google-usm}
Y.~Zhang, W.~Han, J.~Qin, Y.~Wang, A.~Bapna \emph{et~al.}, ``Google {USM}: Scaling automatic speech recognition beyond 100 languages,'' \emph{arXiv preprint arXiv:2303.01037}, 2023.

\bibitem{meta-mms}
V.~Pratap, A.~Tjandra, B.~Shi, P.~Tomasello, A.~Babu \emph{et~al.}, ``Scaling speech technology to 1,000+ languages,'' \emph{Journal of Machine Learning Research}, vol.~25, no.~97, pp. 1--52, 2024.

\bibitem{nv-canary}
K.~C. Puvvada, P.~Żelasko, H.~Huang, O.~Hrinchuk, N.~R. Koluguri \emph{et~al.}, ``Less is more: Accurate speech recognition \& translation without web-scale data,'' in \emph{Proc. Interspeech}, 2024.

\bibitem{peng2024owsmctc}
Y.~Peng, Y.~Sudo, M.~Shakeel, and S.~Watanabe, ``{OWSM}-{CTC}: An open encoder-only speech foundation model for speech recognition, translation, and language identification,'' in \emph{Proc. ACL}, 2024, pp. 10\,192--10\,209.

\bibitem{peng2024owsm}
Y.~Peng, J.~Tian, W.~Chen, S.~Arora, B.~Yan \emph{et~al.}, ``{OWSM} v3.1: Better and faster open {Whisper}-style speech models based on e-branchformer,'' in \emph{Proc. Interspeech}, 2024, pp. 352--356.

\bibitem{li-etal-2024-cb}
Y.~Li, Y.~Li, M.~Zhang, C.~Su, J.~Yu \emph{et~al.}, ``{CB}-{Whisper}: Contextual biasing {Whisper} using open-vocabulary keyword-spotting,'' in \emph{Proc. LREC-COLING}, 2024, pp. 2941--2946.

\bibitem{sun23e_interspeech}
G.~Sun, X.~Zheng, C.~Zhang, and P.~C. Woodland, ``Can contextual biasing remain effective with {Whisper} and {GPT}-2?'' in \emph{Proc. Interspeech}, 2023, pp. 1289--1293.

\bibitem{monteiro24_interspeech}
R.~Monteiro, ``Adding user feedback to enhance {CB}-{Whisper},'' in \emph{Interspeech 2024}, 2024, pp. 347--351.

\bibitem{10800265}
V.~Lall and Y.~Liu, ``Contextual biasing to improve domain-specific custom vocabulary audio transcription without explicit fine-tuning of {Whisper} model,'' in \emph{2024 7th International Conference on Machine Learning and Natural Language Processing (MLNLP)}, 2024, pp. 1--6.

\bibitem{li24c_interspeech}
Y.~Li, M.~Zhang, C.~Su, Y.~Li, X.~Qiao \emph{et~al.}, ``A multitask training approach to enhance {Whisper} with open-vocabulary keyword spotting,'' in \emph{Proc. Interspeech}, 2024, pp. 1260--1264.

\bibitem{chu2023qwen}
Y.~Chu, J.~Xu, X.~Zhou, Q.~Yang, S.~Zhang \emph{et~al.}, ``Qwen-audio: Advancing universal audio understanding via unified large-scale audio-language models,'' \emph{arXiv preprint arXiv:2311.07919}, 2023.

\bibitem{tang2024salmonn}
C.~Tang, W.~Yu, G.~Sun, X.~Chen, T.~Tan \emph{et~al.}, ``{SALMONN}: Towards generic hearing abilities for large language models,'' in \emph{Proc. ICLR}, 2024.

\bibitem{ma2024embarrassingly}
Z.~Ma, G.~Yang, Y.~Yang, Z.~Gao, J.~Wang \emph{et~al.}, ``An embarrassingly simple approach for {LLM} with strong {ASR} capacity,'' \emph{arXiv preprint arXiv:2402.08846}, 2024.

\bibitem{peng2024voicetextblender}
Y.~Peng, K.~C. Puvvada, Z.~Chen, P.~Zelasko, H.~Huang \emph{et~al.}, ``Voice{T}ext{B}lender: Augmenting large language models with speech capabilities via single-stage joint speech-text supervised fine-tuning,'' \emph{arXiv preprint arXiv:2410.17485}, 2024.

\bibitem{gong24b_interspeech}
X.~Gong, A.~Lv, Z.~Wang, and Y.~Qian, ``Contextual biasing speech recognition in speech-enhanced large language model,'' in \emph{Proc. Interspeech}, 2024, pp. 257--261.

\bibitem{higuchi2023harnessing}
Y.~Higuchi, T.~Ogawa, and T.~Kobayashi, ``Harnessing the zero-shot power of instruction-tuned large language model in end-to-end speech recognition,'' \emph{arXiv preprint arXiv:2309.10524}, 2023.

\bibitem{deepcontext2018}
G.~Pundak, T.~N. Sainath, R.~Prabhavalkar, A.~Kannan, and D.~Zhao, ``Deep context: End-to-end contextual speech recognition,'' in \emph{Proc. SLT}, 2018, pp. 418--425.

\bibitem{Jain2020ContextualRF}
M.~Jain, G.~Keren, J.~Mahadeokar, and Y.~Saraf, ``Contextual {RNN}-{T} for open domain asr,'' in \emph{Proc. Interspeech}, 2020, pp. 11--15.

\bibitem{huber2021instant}
C.~Huber, J.~Hussain, S.~St{\"u}ker, and A.~Waibel, ``Instant one-shot word-learning for context-specific neural sequence-to-sequence speech recognition,'' in \emph{Proc. ASRU}, 2021, pp. 1--7.

\bibitem{sudo2023retraining}
Y.~Sudo, K.~Hata, and K.~Nakadai, ``Retraining-free customized {ASR} for enharmonic words based on a named-entity-aware model and phoneme similarity estimation,'' in \emph{Proc. Interspeech}, 2023, pp. 3312--3316.

\bibitem{huang2023contextualized}
K.~Huang, A.~Zhang, Z.~Yang, P.~Guo, B.~Mu \emph{et~al.}, ``{Contextualized End-to-End Speech Recognition with Contextual Phrase Prediction Network},'' in \emph{Proc. Interspeech}, 2023, pp. 4933--4937.

\bibitem{zhou2023copyne}
S.~Zhou, Z.~Li, Y.~Hong, M.~Zhang, Z.~Wang, and B.~Huai, ``Copyne: Better contextual {asr} by copying named entities,'' \emph{arXiv preprint arXiv:2305.12839}, 2023.

\bibitem{panayotov2015librispeech}
V.~Panayotov, G.~Chen, D.~Povey, and S.~Khudanpur, ``Librispeech: an {ASR} corpus based on public domain audio books,'' in \emph{Proc. ICASSP}, 2015, pp. 5206--5210.

\bibitem{shakeel2024_bias}
M.~Shakeel, Y.~Sudo, Y.~Peng, and S.~Watanabe, ``Contextualized end-to-end automatic speech recognition with intermediate biasing loss,'' in \emph{Proc. Interspeech}, 2024, pp. 3909--3913.

\bibitem{csj}
K.~Maekawa, ``Corpus of spontaneous {Japanese}: Its design and evaluation,'' in \emph{ISCA \& IEEE Workshop on Spontaneous Speech Processing and Recognition}, 2003.

\bibitem{futami2023phoneme}
H.~Futami, E.~Tsunoo, Y.~Kashiwagi, H.~Ogawa, S.~Arora, and S.~Watanabe, ``Phoneme-aware encoding for prefix-tree-based contextual {ASR},'' in \emph{Proc. ICASSP}, 2024.

\bibitem{nakagome24_interspeech}
Y.~Nakagome and M.~Hentschel, ``Interbiasing: Boost unseen word recognition through biasing intermediate predictions,'' in \emph{Proc. Interspeech}, 2024, pp. 207--211.

\bibitem{promptasr2024}
X.~Yang, W.~Kang, Z.~Yao, Y.~Yang, L.~Guo \emph{et~al.}, ``Prompt{ASR} for contextualized {ASR} with controllable style,'' in \emph{Proc. ICASSP}, 2024, pp. 10\,536--10\,540.

\bibitem{sudo2024_bias}
Y.~Sudo, M.~Shakeel, Y.~Fukumoto, Y.~Peng, and S.~Watanabe, ``Contextualized automatic speech recognition with attention-based bias phrase boosted beam search,'' in \emph{Proc. ICASSP}, 2024, pp. 10\,896--10\,900.

\bibitem{qiu2023improving}
J.~Qiu, L.~Huang, B.~Li, J.~Zhang, L.~Lu, and Z.~Ma, ``Improving large-scale deep biasing with phoneme features and text-only data in streaming transducer,'' in \emph{Proc. ASRU}, 2023, pp. 1--8.

\bibitem{Le2021ContextualizedSE}
D.~Le, M.~Jain, G.~Keren, S.~Kim \emph{et~al.}, ``Contextualized streaming end-to-end speech recognition with trie-based deep biasing and shallow fusion,'' in \emph{Proc. Interspeech}, 2021, pp. 1772--1776.

\bibitem{sudo2024contextualized}
Y.~Sudo, Y.~Fukumoto, M.~Shakeel, Y.~Peng, and S.~Watanabe, ``Contextualized automatic speech recognition with dynamic vocabulary,'' in \emph{Proc. SLT}, 2024, pp. 78--85.

\bibitem{watanabe2017hybrid}
S.~Watanabe, T.~Hori, S.~Kim, J.~R. Hershey, and T.~Hayashi, ``Hybrid ctc/attention architecture for end-to-end speech recognition,'' \emph{IEEE Journal of Selected Topics in Signal Processing}, vol.~11, no.~8, pp. 1240--1253, 2017.

\bibitem{sudo23c_interspeech}
Y.~Sudo, M.~Shakeel, Y.~Peng, and S.~Watanabe, ``Time-synchronous one-pass beam search for parallel online and offline transducers with dynamic block training,'' in \emph{Proc. Interspeech}, 2023, pp. 4479--4483.

\bibitem{sudo20244d}
Y.~Sudo, M.~Shakeel, Y.~Fukumoto, B.~Yan, J.~Shi, Y.~Peng, and S.~Watanabe, ``Joint beam search integrating ctc, attention, and transducer decoders,'' \emph{IEEE Transactions on Audio, Speech and Language Processing}, vol.~33, pp. 598--612, 2025.

\bibitem{e-branchformer}
K.~Kim, F.~Wu, Y.~Peng \emph{et~al.}, ``{E-Branchformer}: Branchformer with enhanced merging for speech recognition,'' in \emph{Proc. SLT}, 2023, pp. 84--91.

\bibitem{NIPS2017_3f5ee243}
A.~Vaswani, N.~Shazeer, N.~Parmar, J.~Uszkoreit, L.~Jones, A.~N. Gomez, L.~u. Kaiser, and I.~Polosukhin, ``Attention is all you need,'' in \emph{Proc. NeurIPS}, 2017, pp. 5998--6008.

\bibitem{KingmaB14}
D.~P. Kingma and J.~Ba, ``Adam: A method for stochastic optimization,'' in \emph{Proc. ICLR}, 2015.

\bibitem{espnet}
S.~Watanabe, T.~Hori, S.~Karita, T.~Hayashi, J.~Nishitoba \emph{et~al.}, ``{ESPnet}: End-to-end speech processing toolkit,'' in \emph{Proc. Interspeech}, 2018, pp. 2207--2211.

\end{thebibliography}

\end{document}